\documentclass[pre,twocolumn,showpacs,preprintnumbers,floatfix]{revtex4-1}

\usepackage{etex}
\usepackage{ifpdf}
\usepackage{hyperref}
\usepackage{dcolumn}
\usepackage{url}
\usepackage{amsmath}
\usepackage{amscd}
\usepackage{amsfonts}
\usepackage{amssymb}
\usepackage{amsthm}
\usepackage{xfrac}
\usepackage{braket}
\usepackage{bm}   
\usepackage{bbm}
\usepackage{verbatim}
\usepackage{stmaryrd}
\usepackage{xcolor}
\usepackage{setspace}

\usepackage[resetlabels]{multibib}
\newcites{supp}{Supplementary Citations}
\bibliographystylesupp{unsrt}

\newcommand{\eM}     {\mbox{$\epsilon$-machine}}
\newcommand{\eMs}    {\mbox{$\epsilon$-machines}}

\newcommand{\eT}     {\mbox{$\epsilon$-transducer}}
\newcommand{\eTs}    {\mbox{$\epsilon$-transducers}}
\newcommand{\ET}     {\mbox{$\epsilon$-Transducer}}

\newcommand{\ETs}    {\mbox{$\epsilon$-Transducers}}

\newcommand{\MeasAlphabet}  {\mathcal{A}}
\newcommand{\MeasSymbol}   { {X} }
\newcommand{\meassymbol}   { {x} }

\newcommand{\CausalState}   { \mathcal{S} }

\newcommand{\causalstate}   { \sigma }
\newcommand{\CausalStateSet}    { \boldsymbol{\CausalState} }
\newcommand{\AlternateState}    { \mathcal{R} }

\newcommand{\AlternateStateSet} { \boldsymbol{\AlternateState} }

\newcommand{\Prob}      {\Pr} 

\newcommand{\Cmu}       {C_\mu}
\newcommand{\hmu}       {h_\mu}
\newcommand{\EE}        {{\bf E}}

\newcommand{\ProcessAlphabet}   {\MeasAlphabet}

\newcommand{\forward}{+}
\newcommand{\reverse}{-}
\newcommand{\forwardreverse}{\pm} 

\newcommand{\FutureCausalState} { {\CausalState}^{\forward} }

\newcommand{\PastCausalState}   { {\CausalState}^{\reverse} }

\newcommand{\lastindex}[2]{
  \edef\tempa{0}
  \edef\tempb{#2}
  \ifx\tempa\tempb
    \edef\tempc{#1}
  \else
    \edef\tempa{0}
    \edef\tempb{#1}
    \ifx\tempa\tempb
      \edef\tempc{#2}
    \else
      \edef\tempc{#1+#2}
    \fi
  \fi
  \tempc
}

\newcommand{\I}{\mathbf{I}}

\newcommand{\CSjoint}[1][,]{
   \edef\tempa{:}
   \edef\tempb{#1}
   \ifx\tempa\tempb
      \ensuremath{\FutureCausalState\!#1\PastCausalState}
   \else
      \ensuremath{\FutureCausalState#1\PastCausalState}
   \fi
}

\newif\ifpm
\edef\tempa{\forwardreverse}
\edef\tempb{\pm}
\ifx\tempa\tempb
   \pmfalse
\else
   \pmtrue
\fi

\renewcommand{\H}{\operatorname{H}}
\renewcommand{\I}{\operatorname{I}}

\def\clap#1{\hbox to 0pt{\hss#1\hss}}

\newenvironment{itemize*}%
  {\begin{itemize}%
    \setlength{\itemsep}{0pt}%
    \setlength{\parskip}{0pt}%
    \setlength{\topsep}{0pt}%
    \setlength{\partopsep}{0pt}%
    \setlength{\parsep}{0pt}}%
  {\end{itemize}}

\newenvironment{enumerate*}%
  {\begin{enumerate}%
    \setlength{\itemsep}{0pt}%
    \setlength{\parskip}{0pt}%
    \setlength{\topsep}{0pt}%
    \setlength{\partopsep}{0pt}%
    \setlength{\parsep}{0pt}}%
  {\end{enumerate}}

\begin{document}

\title{Demon Dynamics:\\
Deterministic Chaos, the Szilard Map, and\\
the Intelligence of Thermodynamic Systems}

\author{Alexander B. Boyd}
\email{abboyd@ucdavis.edu}

\author{James P. Crutchfield}
\email{chaos@ucdavis.edu}

\affiliation{Complexity Sciences Center and Department of Physics,\\
University of California at Davis, One Shields Avenue, Davis, CA 95616}

\date{\today}
\bibliographystyle{unsrt}

\begin{abstract}
We introduce a deterministic chaotic system---the Szilard Map---that
encapsulates the measurement, control, and erasure protocol by which Maxwellian
Demons extract work from a heat reservoir. Implementing the Demon's control
function in a dynamical embodiment, our construction symmetrizes Demon and
thermodynamic system, allowing one to explore their functionality and recover
the fundamental trade-off between the thermodynamic costs of dissipation due to
measurement and due to erasure. The map's degree of chaos---captured by the
Kolmogorov-Sinai entropy---is the rate of energy extraction from the heat bath.
Moreover, an engine's statistical complexity quantifies the minimum necessary
system memory for it to function. In this way, dynamical instability in the
control protocol plays an essential and constructive role in intelligent
thermodynamic systems.


\vspace{0.2in}

\end{abstract}

\preprint{Santa Fe Institute Working Paper 15-06-XXX}
\preprint{arxiv.org:1506.XXXX [cond-mat.stat-mech]}

\maketitle


\setstretch{1.1}

\newcommand{\Abet}{\ProcessAlphabet}
\newcommand{\MS}{\MeasSymbol}
\newcommand{\ms}{\meassymbol}
\newcommand{\SSet}{\CausalStateSet}
\newcommand{\St}{\CausalState}
\newcommand{\st}{\causalstate}
\newcommand{\MxSt}{\AlternateState}
\newcommand{\MxSSet}{\AlternateStateSet}
\newcommand{\mxst}{\eta}
\newcommand{\mxstt}[1]{\eta_{#1}}
\newcommand{\StartMS}{\bra{\delta_\pi}}

\newcommand{\MapSym} { \mathcal{T} }
\newcommand{\TM} { \MapSym_\mathrm{M} }
\newcommand{\TMeas} { \MapSym_\mathrm{Measure} }
\newcommand{\TC} { \MapSym_\mathrm{C} }
\newcommand{\TTherm} { \MapSym_\mathrm{Thermodynamics} }
\newcommand{\TT} { \MapSym_\mathrm{T} }
\newcommand{\TCon} { \MapSym_\mathrm{Control} }
\newcommand{\TE} { \MapSym_\mathrm{E} }
\newcommand{\TEr} { \MapSym_\mathrm{Erase} }
\newcommand{\TIE} { \MapSym_\mathrm{IE} }
\newcommand{\TB} { \MapSym_\mathrm{B} }
\newcommand{\TBake} { \MapSym_\mathrm{Baker} }
\newcommand{\TS} { \MapSym_\mathrm{S} }
\newcommand{\TSzil} { \MapSym_\mathrm{Szilard} }
\newcommand{\TSzilComposite} { \widehat{\MapSym}_\mathrm{Szilard} }

\newcommand{\QM} { Q_\mathrm{measure} }
\newcommand{\QE} { Q_\mathrm{erase} }
\newcommand{\QC} { Q_\mathrm{control} }
\newcommand{\QD} { Q_\mathrm{diss} }

\newcommand{\IESymbol} { Z }
\newcommand{\IEPast}   { \smash{\overleftarrow  {\IESymbol}} }
\newcommand{\IEFuture} { \smash{\overrightarrow {\IESymbol}} }
\newcommand{\IEEE}     { \EE^{\text{IE}} }
\newcommand{\DEE}      { \EE^{\text{Demon}} }
\newcommand{\SUSEE}    { \EE^{\text{SUS}} }
\newcommand{\IECmu}    { \Cmu^{\text{IE}} }
\newcommand{\DCmu}     { \Cmu^{\text{Demon}} }
\newcommand{\SUSCmu}   { \Cmu^{\text{SUS}} }
\newcommand{\IEhmu}    { \hmu^{\text{IE}} }
\newcommand{\Dhmu}     { \hmu^{\text{Demon}} }
\newcommand{\SUShmu}   { \hmu^{\text{SUS}} }


Synthetic nanoscale machines \cite{Ruek00a,Fenn03a,Zhon03a,Chen06a}, like their
macromolecular biological counterparts \cite{Juli97a,Bust05a,Dunn15a}, perform
tasks that involve the simultaneous manipulation of energy, information, and
matter. In this they are information engines---systems with two inextricably
intertwined characters. The first aspect, call it ``physical'', is the one in
which the system---seen embedded in a material substrate---is driven by,
manipulates, stores, and dissipates energy. The second aspect, call it
``informational'', is the one in which the system---seen in terms of its
spatial and temporal organization---generates, stores, loses, and transforms
information.  Information engines operate by synergistically balancing both
aspects to support a given functionality, such as extracting work from a heat
reservoir.

This is remarkable behavior. Though we can sometimes identify it---in a motor
protein hauling nutrients across a cell's microtubule highways \cite{Juli97a},
in how a quantum dot transistor modulates current under the influence of an
evanescent wave function \cite{Zhua98a,Hets13a}---it is not well understood.
Understanding calls on a thermodynamics of nanoscale systems that operate far
out of equilibrium and on a physics of information that quantitatively
identifies organization and function, neither of which has been fully articulated. However, recent theoretical and experimental breakthroughs
\cite{Bust05a,Toya10a,Beru12a,Klag13a,Dunn15a} hint that we may be close to a
synthesis which not only provides understanding but predicts quantitative,
measurable functionalities.

We define an \emph{information engine} as a system that performs information
processing as it undergoes controlled thermodynamic transformations. We show
that information engines are chaotic dynamical systems. Building this bridge to
dynamical systems theory allows us to employ its powerful tools to analyze an
engine's complex, nonlinear behavior. This includes a thorough informational
and structural analysis that leads to a measure of thermodynamic system
intelligence.

By way of concretely illustrating the theory, we introduce an explicit
implementation of Szilard's Engine \cite{Szil29a} as an iterated composite map
of the unit square that is a deterministic, but chaotic dynamical system. The
result is a particularly simple and constructive view of the energetics and
computation embedded in controlled nonlinear thermodynamical systems. That
simplicity, however, gives a solid base for designing and analyzing real
information engines. We end giving a concise statement of the general theory
and applications.

\paragraph*{Background}

The Szilard Engine is an ideal Maxwellian Demon for examining the role of
information processing in the Second Law \cite{Szil29a}. The engine consists of
three components: a controller (the Demon), a thermodynamic system (a particle
in a box), and a heat bath that keeps both thermalized to a temperature $T$. It
operates by a simple mechanism of a repeating three-step cycle of measurement,
control, and erasure. During measurement, a barrier is inserted midway in the
box, constraining the particle either to box's left or right half, and the
Demon memory changes to reflect on which side the particle is. In the
thermodynamic control step, the Demon uses that knowledge to
allow the particle to push the barrier, extracting $\int P~dV = k_B T \ln 2$
work from the thermal bath. (Supplementary Materials review this and related
thermodynamic calculations.) In the erasure step, the Demon resets its finite
memory to a default state, so that it can perform measurement again. The
periodic \emph{protocol} cycle of measurement, control, and erasure repeats
endlessly and deterministically. The net result being the extraction of work
from the bath balanced by entropy created by changes in the Demon's memory.

Rather than seeing the Demon and box as separate, though, we view it---an
information engine---as the direct product of thermodynamic system and Demon
memory \footnote{Cf. the schematic diagram in Ref. \cite[Fig. 12]{Benn82a}, a
more complicated product system, used to argue that only erasure during
logically irreversible operations dissipates energy. This, however, is a
restricted case.}. Though we follow Szilard closely, he did not specify the
Demon's physical embodiment. Critically, we choose the Demon's memory to be
another spatial dimension of a particle in a box. Thus, we see the joint system
as a single particle in a two-dimensional box, where one axis represents the
thermodynamic System Under Study (SUS)---the original particle in a box---and
the other axis represents the Demon memory. We now describe a deterministic
protocol that implements the Szilard Engine, evolving a particle ensemble
over the joint state space.

\begin{figure*}[t]
\begin{center}
\includegraphics[width=2.1\columnwidth]{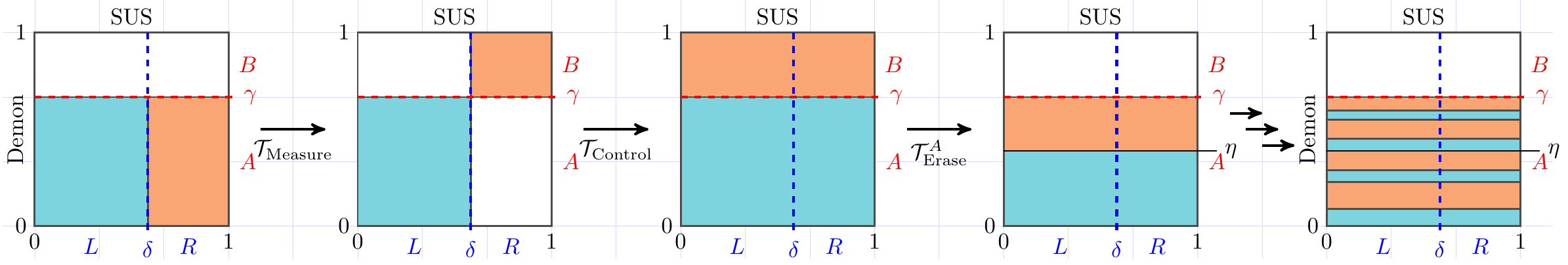}\\
\end{center}
\caption{Szilard Engine as a deterministic dynamical system: the
  Szilard Map $\TSzil = \TEr^A \circ \TCon \circ \TMeas$. Regions left
  and right of $\delta$ colored to aid tracking particle location.
  Rightmost: action of $\TSzil^3$ resulting in self-similar (fractal)
  structure in density $\widehat{\rho}$; uniform $\widehat{\rho}$
  requires $\eta = \gamma \delta$. These assume the Demon's reset
  memory state is $A$. (Supplementary Materials include an animation.)
  }
\label{fig:SMap}
\end{figure*}

\paragraph*{A Dynamical Engine}
The Szilard Engine's measurement-control-erasure barrier-sliding protocol is
equivalent to a discrete-time two-dimensional map from unit square
$\mathbb{I}^2 = [0,1] \times [0,1]$ to itself. The engine has two kinds of
\emph{mesoscopic states}---states of the particle's location $\big\{L \sim x \in
(0,\delta], R \sim x \in (\delta,1) \big\}$ and states of the Demon's knowledge
$\big\{A \sim y \in (0,\gamma], B \sim y \in (\gamma,1) \big\}$ of the
location---that partition the joint states $(x,y) \in \mathbb{I}^2$.

The protocol cycle translates into a composite map $\TSzil = \TE \circ \TC
\circ \TM$ of $\mathbb{I}^2$; one map for each engine step; see Fig.
\ref{fig:SMap}(a). As they operate, they take the joint state space from one
stage to another around the cycle:
\paragraph*{Measurement:} To correlate Demon memory with particle
location $\TM$ takes the $A\otimes L$ and the $B \otimes L$
mesostates to themselves, the $A \otimes R$ mesostate to $B \otimes R$, and $B
\otimes R$ to $A \otimes R$:
\begin{align*}
\TM(x,y) =
\begin{cases}
(x,y) & x < \delta, y < \gamma
  \text{ or } x < \delta, y \geq \gamma ~,\\
  \left(x, \gamma + y \frac{1-\gamma}{\gamma}\right) &
  x \geq \delta, y \leq \gamma ~,\\
  \left( x, \gamma \frac{y-\gamma}{1-\gamma} \right) &
  x \geq \delta, y > \gamma
  ~.
\end{cases}
\end{align*}
\paragraph*{Thermodynamic control:} To extract energy from the bath $\TC$ expands both the $A$ and $B$ Demon
memory mesostates over the SUS's whole interval:
\vspace{-0.1in}
\begin{align*}
\TC(x,y) =
\begin{cases}
  (\frac{x}{\delta},y) & x < \delta ~, \\
  \left(\frac{x-\delta}{1-\delta}, y\right) & x \geq \delta
  ~.
\end{cases}
\end{align*}
\vspace{-0.1in}
\paragraph*{Erasure:} $\TE$ maps both the $A$ and $B$ Demon memory mesostates
back to a selected Demon memory reset mesostate. If that reset state is $A$,
then the mapping is:
\vspace{-0.1in}
\begin{align*}
\TE^A(x,y) =
\begin{cases}
  (x,y\delta) & y < \gamma ~, \\
  \left(x,\delta\gamma+\frac{y-\gamma}{1-\gamma}\gamma(1-\delta) \right)
  & y \geq \gamma ~.
\end{cases}
\end{align*}
\vspace{-0.1in}

This explicit construction establishes that Szilard's Engine is a deterministic
dynamical system whose component maps are thermodynamic transformations---a
\emph{piecewise thermodynamical system}. The mapping $\TSzil$ means we can
avail ourselves of the analytical tools of dynamical systems theory
\cite{Laso85a,Dorf99a} to analyze the Szilard Engine mechanisms. This perforce
suggests a number of more refined and quantitative questions about the dynamics
ranging from the structural role of the stable and unstable submanifolds in
supporting information and thermodynamic processing and the existence of
asymptotic invariant measures to measures of information generation, storage,
and intelligence.

As typically done to establish a known initial state for any engineered computing device, we initialize the system first using $\TE$. The result is that $\TSzil$ becomes the well known asymmetric Baker's Map $\TBake^A (x,y)  = \TC \circ \TM \circ \TE^A$:
\vspace{-0.1in}
\begin{align*}
\TBake^A (x,y) = 
 & \begin{cases}
 (\frac{x}{\delta},\delta y)
 & x < \delta ~,\\
 \left( \frac{x-\delta}{1-\delta}, \delta + y (1-\delta) \right)
 & x \geq \delta ~.
\end{cases}
\end{align*}
The familiar stretching and folding action of $\TBake$ is shown in
Fig.~\ref{fig:SMapComposite}(b) of the Supplementary Materials. Being a Baker's
map, it is immediately clear that the Szilard Engine dynamics are chaotic
\cite{Laso85a,Dorf99a}.

While the overall composite map $\TSzil$ is important, considering its
complete-cycle behavior alone misses much. What is key are the component maps
that nominally control a thermodynamic system, with each step corresponding to
a different thermodynamic transformation. We now analyze the dynamics to see
how the component maps contribute to information processing and thermodynamics.
(Supplementary Materials give calculational details.)

\paragraph*{Dynamical Systems Analysis}
What does chaos in the Szilard Engine mean? The joint system generates
information---the information about particle position that the Demon must keep
repeatedly measuring to stay synchronized to the SUS and so extract energy from
the bath. On the one hand, it is generated by the heat bath through state-space
expansion during $\TC$. And, on the other, it is stored by the Demon
(temporarily) and must be erased during $\TE$. The latter's construction makes
clear that it, dynamically, contracts state-space and so is locally dissipative.

With explicit equations of motion in hand, one can directly check, by
calculating the Jacobian $\partial_{xy} \TBake$, that the map is globally area
preserving.  Moreover, the invariant distribution $\widehat{\rho}$ can be
determined from the Frobenious-Perron operator \cite{Laso85a,Dorf99a}:

\vspace{-0.1in}
\begin{align*}
\widehat{\rho} (x^\prime,y^\prime)
  = \int_{\mathbb{I}^2} dx dy ~
  \delta \big( (x^\prime,y^\prime) - \TBake (x,y) \big)
  \widehat{\rho} (x,y)
  ~.
\end{align*}
($\delta(\cdot)$ here, and only here, is the Dirac delta-function.)
Calculation shows that $\widehat{\rho}$ has full support on the unit square $\mathbb{I}^2$ and so its fractal dimension is $d_f = 2$ for all $\delta, \gamma \in (0,1)$. The particle density is uniform when, during $\TE$, the Demon's memory mesostate partition falls at $\eta = \gamma \delta$. However,
when $\eta \neq \gamma \delta$, the density is not uniform, which is reflected
in $\widehat{\rho}$'s information dimension $d_I < d_f$,
\cite[Chs. 11-12]{Dorf99a}. This corresponds to
changing the efficiency of the Demon's information extraction, which we see is
reflected in the difference $d_f - d_I$.

The Szilard Map Jacobian also determines its local linearization and so one can
easily calculate the spectrum of Lyapunov characteristic exponents (LCEs) for
the overall cycle and so realize the contribution of each protocol step.
This gives insight into the directions (submanifolds) of
stability (convergence) and instability (divergence). There are two LCEs: one
positive $\lambda^+  = \H(\delta)$ and one negative $\lambda^- = - \H(\delta)$,
where $\H(\delta)$ is the (base $2$) binary entropy function \cite{Cove06a}.
(See Supplementary Material for details.)
Note that
energy conservation ($\TSzil$'s area preservation) is reflected in the exact
balance of instability and stability: $\lambda^+ + \lambda^- = 0$. The unstable
manifolds (parallel the $x$-axis) support the mechanism that amplifies small
fluctuations from the heat bath to macroscopic scale during energy extraction
($\TC$). The stable manifolds (parallel the $y$-axis) are the mechanism that
dissipates energy into the ambient heat bath, during erasure ($\TE$).

The overall information production rate is given by $\TSzil$'s Kolmogorov-Sinai
entropy $\hmu$ \cite{Sina59}. For the Szilard Engine, given the well
behaved nature of $\widehat{\rho}$, $\hmu = \sum_{\lambda > 0} \lambda =
\lambda^+$ by Pesin's Theorem \cite{Dorf99a}. (That is, $\hmu = \H(\delta)$,
directly verified shortly.) This measures the flow of information from the SUS
into the Demon: information harvested from the bath and used by the Demon to
convert thermal energy into work. Simply stated, the degree of chaos determines
the rate of energy extraction from the bath.

\begin{figure}[htp]
\begin{center}
\includegraphics[width=0.7\columnwidth]{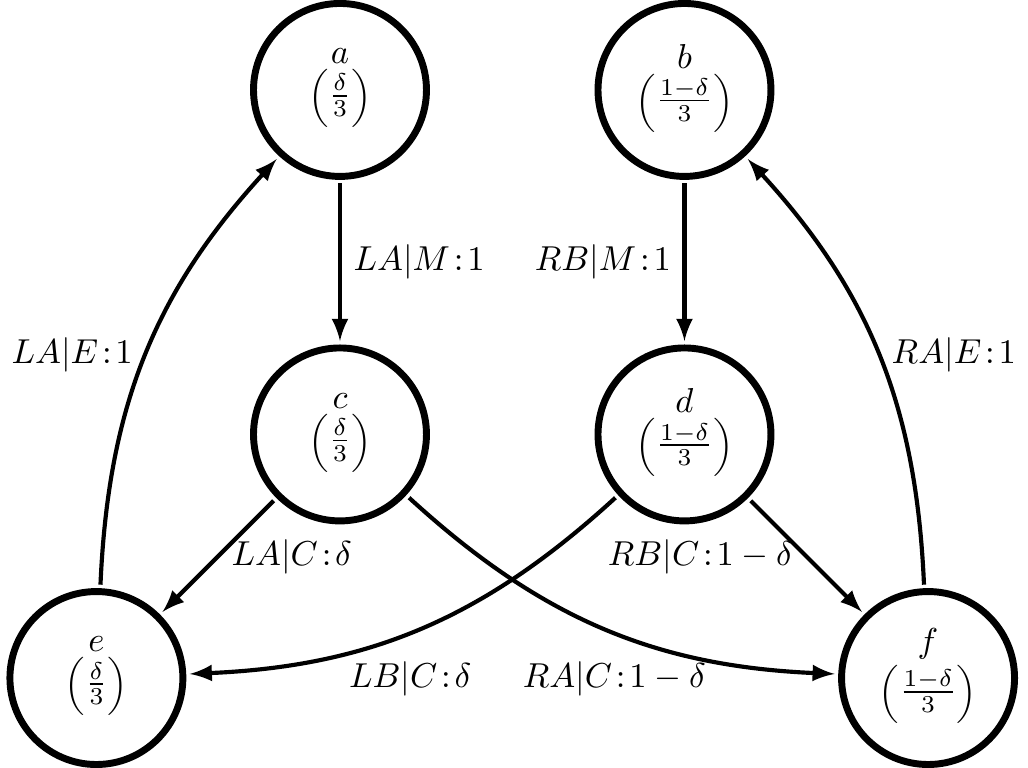}
\includegraphics[width=0.2\columnwidth]{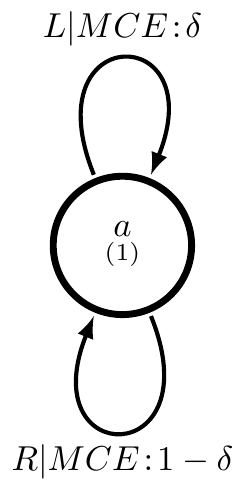} \\
\hspace{0.85in} (a) \hspace{1.25in} (b)
\end{center}
\caption{\ETs\ for the symbolic dynamics of the Szilard information engine
  from the Markov partition of its joint state space:
  (a) The \eT\ for $\TSzil$ that reads in the periodic control signal
  for measure ($M$), control (C), and erase (E).
  (b) $\TSzil$ single-state \eT: Memoryless over the full measure-control-erase
  protocol. Transitions $\beta | \alpha : p$ denote reading protocol symbol
  $\alpha$, taking the transition with probability $p$, emitting symbol $\beta$.
  Asymptotic state probabilities are given in parentheses underneath
  state names.
  }
\label{fig:SzilardEM}
\end{figure}

\paragraph*{Computational Mechanics Analysis}
The Demon memory and particle location mesostates form Markov partitions for
the Szilard Map dynamics \cite[Chs. 7,9]{Dorf99a}: tracking sequences of symbols in
$\{A,B\}$ or in $\{R,L\}$ (or all four pairs $\{AR, AL, BR, BL\}$) leads to a
symbolic dynamics that captures all of the joint system's information
processing behavior. We now use this fact to analyze the various kinds of
information processing and introduce a way to measure the Demon's
``intelligence'' or, more appropriately, that of the entire engine. We do this
by calculating computational mechanics' \eMs\ and \eTs\ from the engine's
symbolic dynamics \cite{Crut12a,Barn13a}. The overall engine transducer is
shown in Fig. \ref{fig:SzilardEM}(a), that for the SUS particle dynamics in
Fig. \ref{fig:SzilardEMSupp}(a), and for the Demon memory dynamics in Fig.
\ref{fig:SzilardEMSupp}(b).

In addition to explicitly expressing the effective mechanisms that support
information processing, \eMs\ allow us to quantify the effects of measurement,
control, and erasure. The engine's Kolmogorov-Sinai entropy $\hmu$ can be
calculated directly from the \eMs's causal-state averaged transition
uncertainty.  To quantitatively measure the minimal required memory---a key
component of ``intelligence''---for the information engine functioning, we
employ the \eM's statistical complexity $\Cmu = \H[\Prob(\causalstate)]$, where
$\causalstate \in \CausalStateSet$ are the system's causal states
\cite{Crut12a} and $\H[\cdot]$ is the Shannon information \cite{Cove06a}.

It is important to emphasize an aspect of the information engine
\eM\ construction: It is stage-dependent in that, to fully capture the component
operations and their thermodynamic effects, the individual maps must to taken
into account. This observation should be contrasted with the symbolic dynamics
and particle position \eM\ for the \emph{overall} Szilard map in its Baker's
map form. The resulting process arises from stroboscopically observing the
behavior after driving the engine with the three-symbol word $MCE$. As an
example, the particle position process's \eM\ is shown in Fig.
\ref{fig:SzilardEM}(b); it is a biased coin---a single-state \eM\ with no
memory: $\Cmu(\TBake) = 0$. This is as it should be: The overall cycle must
return to the same state storing no memory of previous cycles.

Computational mechanics analysis shows that, over the three-step cycle, the
Engine has an entropy rate of $\hmu = \H(\delta)$ as seen above (or
$\H(\delta)/3$ per map step) and a statistical complexity of $\Cmu = \log_2 3 +
\H(\delta)$. (See Supplementary Materials for details, including analysis of
SUS and Demon subsystems.) How predictable is the Engine's operation? The
information in its future predictable from its past is given by the
\emph{excess entropy}: $\EE = \I[\IEPast;\IEFuture]$, where $\IEPast$ is the
past and $\IEFuture$ is the future of the joint process over random variable
$\IESymbol_t \in \{A,B\} \otimes \{R,L\}$. We see that the machine in Fig.
\ref{fig:SzilardEM}(a), driven by the protocol, is counifilar and so $\EE =
\Cmu$ \cite{Crut08b}.

\paragraph*{Thermodynamics}
During each protocol step the Engine interacts thermodynamically with the heat
bath. The Supplementary Materials calculate the average heat $\braket{Q}$ and
work $\braket{W}$ flows between the Demon and the bath and between the SUS and
the bath during each step. For the Szilard Engine heat and work are equivalent
and so we discuss only the heat as energies $\braket{Q_{\text{diss}}}$
dissipated to the bath for each interaction. As we will see, although
$\gamma$---the Demon memory partition---did not play a direct role in the
informational properties, it does in the thermodynamics.

The expected heat flow during measurement is $\braket{\QM} = -k_B T (1-\delta)
\ln \left( (1-\gamma) / \gamma \right)$. Since $\gamma \in [0,1]$, the
dissipated heat can be negative or positive. It vanishes at $\gamma = 1/2$.
Negative dissipated heat means that the engine absorbs energy from the heat
bath and, in that case, turns it into work. The work $\int P~dV$ done by the
particle on the barrier is $k_B T \H(\delta) \ln 2$. Thus, the average heat
absorbed by the engine from the heat bath during thermodynamic control is
$\braket{\QC} = - k_B T \H(\delta) \ln 2$, which is maximized when $\delta =
1/2$. During memory erasure the Demon shifts back to its default state, without
affecting the SUS state. The barrier partitioning the Demon's mesostates slides,
compressing the contained particle into the default state $A$, say. The heat
dissipated in this process is $\braket{\QE} = k_B T (1-\delta) \ln \left(
(1-\gamma) / \gamma \right) + k_B T \H(\delta)\ln 2$.

\begin{figure}[htp]
\begin{center}
\includegraphics[width=0.95\columnwidth]{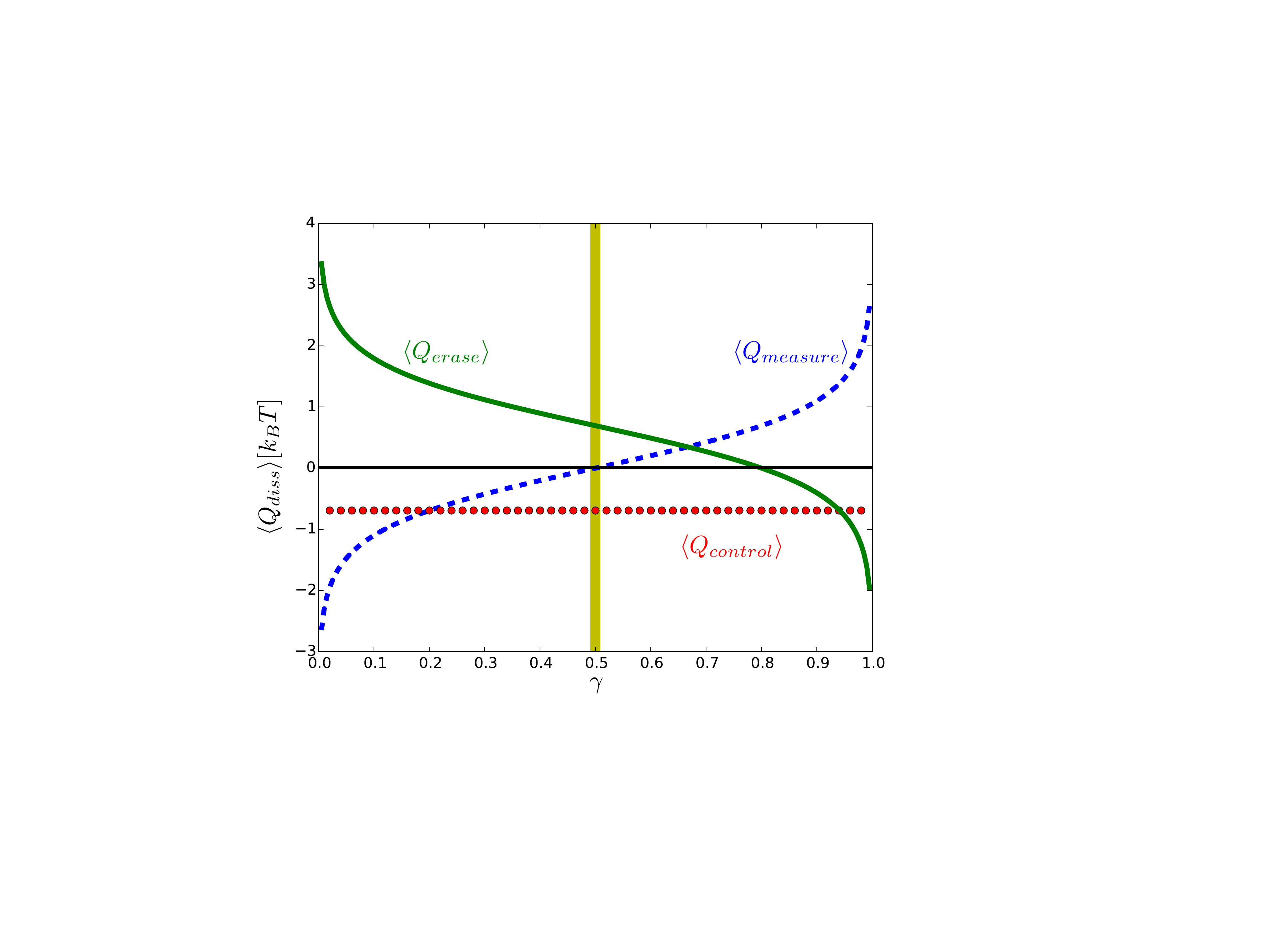}
\end{center}
\caption{Beyond Landauer's Principle: Thermodynamic costs (energy dissipation
  $\QD$) for measurement, control, and erasure in Szilard's information engine
  as a function of $\gamma$ (Demon partition) with SUS barrier at $\delta =
  1/2$. Landauer's Principle applies only at $\gamma = 1/2$ (vertical yellow
  band): measurements are thermodynamically free, erasure costs since heat is
  dissipated as a result of Demon resetting. Costs \emph{exactly flip} at
  $\gamma = 4/5$, though.
  }
\label{fig:ThermoCosts}
\end{figure}

While the heat dissipated during control is independent of $\gamma$, both
measurement and erasure can dissipate any positive or negative amount of heat,
depending on $\gamma$. Notably, for $\gamma > 1/2$, the Szilard Engine violates
Landauer's Principle \cite{Land61a,Benn82a} in that $\braket{\QE}
\leq \ln 2$; in energy units of $k_B T$. Indeed, for $\gamma = 4/5$, erasure is
thermodynamically free and measurement is costly---an anti-Landauer Principle.

Figure \ref{fig:ThermoCosts} illustrates the trade-offs in thermodynamic costs
for each step. They sum to zero and so the Engine respects the Second Law over
the whole range of $\delta$ and $\gamma$. The erasure and measurement steps
together obey the relation:
$\braket{\QE} + \braket{\QM} = k_B T \H(\delta)\ln 2$,
recovering trade-offs noted previously \cite{Shiz95a,Fahn96a,Bark06a,Saga12a}.
That is, the Szilard Engine achieves the lower bounds on energy dissipation
during measurement and erasure. And so, it plays an analogous optimal role in
the conversion of information into work as the Carnot Engine does for optimal
efficiency when converting thermal energy to work.

\paragraph*{Final Remarks}
We leveraged a straightforward observation to give a thorough dynamical
systems, computational mechanics, and thermodynamic analysis of Szilard's
Engine: an information engine's intrinsic computation is supported by the
evolution of its joint state-space distribution and its thermodynamic costs
monitor how those distributional changes couple energetically to its environment.

The Szilard Map construction is straightforward and easy to interpret. For
these reasons, we selected it to illustrate the bridge between thermodynamics,
information theory, and dynamical systems necessary to fully analyze
information engines. The approach generalizes. We can now state \emph{our
central proposal}: (i) an information engine is the dynamic over a joint state
space of a thermodynamic system and a physically embodied controller, (ii) the
causal states of the joint dynamics, formed from the
predictive equivalence classes of system histories, capture its information
processing and emergent organization, (iii) a necessary component of the
engine's effective ``intelligence'', its memory,
is given by its statistical complexity $\Cmu$, (iv) its dissipation is given by
the dynamical system negative LCEs, and (v) the rate of energy extracted from
the heat bath is governed by the Kolmogorov-Sinai entropy $\hmu$.

Sequels use this approach to analyze the information thermodynamics of more
sophisticated engines, including the Mandal-Jarzynski ratchet \cite{Mand012a},
experimental nanoscale information processing devices, and intelligent
macromolecules.

\paragraph*{Supplementary Materials:} Calculational details, further
discussion and interpretation, and animation illustrating
a continuous-time embedding of $\TSzil$.

\paragraph*{Acknowledgments} 
We thank Gavin Crooks, David Feldman, Ryan James, John Mahoney, Sarah Marzen,
Paul Riechers, and Michael Roukes for helpful discussions. Work supported by
the U.S. Army Research Laboratory and the U. S. Army Research Office under
contracts W911NF-13-1-0390 and W911NF-12-1-0234.

\vspace{-0.25in}
\bibliography{chaos}

\begin{thebibliography}{10}

\bibitem{Ruek00a}
T.~Rueckes, K.~Kim, E.~Joselevich, Greg~Y. Tseng, C-L. Cheung, and C.~M.
  Lieber.
\newblock {\em Science}, 289:94--97, 2000.

\bibitem{Fenn03a}
A.~M. Fennimore, T.~D. Yuzvinsky, W.-Q. Han, M.~S. Fuhrer, J.~Cumings, and
  A.~Zettl.
\newblock {\em Nature}, 424:408--410, 2003.

\bibitem{Zhon03a}
Z.~Zhong, D.~Wang, Y.~Cui, M.~W. Bockrath, and C.~M. Lieber.
\newblock {\em Science}, 302:1377--1379, 2003.

\bibitem{Chen06a}
J.~Chen, N.~Jonoska, and G.~Rozenberg, editors.
\newblock {\em Nanotechnology: Science and Computation}, Natural Computing, New
  York, 2006. Springer-Verlag.

\bibitem{Juli97a}
F.~Julicher, A.~Ajdari, and J.~Prost.
\newblock {\em Rev. Mod. Phys.}, 69(4):1269--1281, 1997.

\bibitem{Bust05a}
C.~Bustamante, J.~Liphardt, and F.~Ritort.
\newblock {\em Physics Today}, 58(7):43--48, 2005.

\bibitem{Dunn15a}
A.~R. Dunn and A.~Price.
\newblock {\em Physics Today}, 68(2):27--32, 2015.

\bibitem{Zhua98a}
L.~Zhuang, L.~Guo, and S.~Y. Chou.
\newblock {\em Appl. Phys. Lett.}, 72(10):1205--1207, 1998.

\bibitem{Hets13a}
F.~Hetsch, N.~Zhao, S.~V. Kershaw, and A.~L. Rogach.
\newblock {\em Materials Today}, 16(9):312--325, 2013.

\bibitem{Toya10a}
S.~Toyabe, T.~Sagawa, M.~Ueda, E.~Muneyuki, and M.~Sano.
\newblock {\em Nature Physics}, 6:988--992, 2010.

\bibitem{Beru12a}
A.~Berut, A.~Arakelyan, A.~Petrosyan, S.~Ciliberto, R.~Dillenschneider, and
  E.~Lutz.
\newblock {\em Nature}, 483:187--190, 2012.

\bibitem{Klag13a}
R.~Klages, W.~Just, and C.~Jarzynski, editors.
\newblock Wiley, New York, 2013.

\bibitem{Szil29a}
L.~Szilard.
\newblock {\em Z. Phys.}, 53:840--856, 1929.

\bibitem{Note1}
Cf. the schematic diagram in Ref. \cite [Fig. 12]{Benn82a}, a more complicated
  product system, used to argue that only erasure during logically irreversible
  operations dissipates energy. This, however, is a restricted case.

\bibitem{Laso85a}
A.~Lasota and M.~C. Mackey.
\newblock {\em Probabilistic Properties of Deterministic Systems}.
\newblock Cambridge University press, Cambridge, United Kingdom, 1985.

\bibitem{Dorf99a}
J.~R. Dorfman.
\newblock {\em An Introduction to Chaos in Nonequilibrium Statistical
  Mechanics}.
\newblock Cambridge University Press, Cambridge, United Kingdom, 1999.

\bibitem{Cove06a}
T.~M. Cover and J.~A. Thomas.
\newblock {\em Elements of Information Theory}.
\newblock Wiley-Interscience, New York, second edition, 2006.

\bibitem{Sina59}
Ja.~G. Sinai.
\newblock {\em Dokl. Akad. Nauk. SSSR}, 124:768, 1959.

\bibitem{Crut12a}
J.~P. Crutchfield.
\newblock {\em Nature Physics}, 8(January):17--24, 2012.

\bibitem{Barn13a}
N.~Barnett and J.~P. Crutchfield.
\newblock {\em J. Stat. Phys.}, to appear, 2015.

\bibitem{Crut08b}
C.~J. Ellison, J.~R. Mahoney, and J.~P. Crutchfield.
\newblock {\em J. Stat. Phys.}, 136(6):1005--1034, 2009.

\bibitem{Land61a}
R.~Landauer.
\newblock {\em IBM J. Res. Develop.}, 5(3):183--191, 1961.

\bibitem{Benn82a}
C.~H. Bennett.
\newblock {\em Intl. J. Theo. Phys.}, 21:905, 1982.

\bibitem{Shiz95a}
K.~Shizume.
\newblock {\em Phys. Rev. E}, 52(4):3495--3499, 1995.

\bibitem{Fahn96a}
F.~N. Fahn.
\newblock {\em Found. Physics}, 26(1):71--93, 1996.

\bibitem{Bark06a}
M.~M. Barkeshli.
\newblock {\em arXiv:cond-mat}, 0504323.

\bibitem{Saga12a}
T.~Sagawa.
\newblock {\em Prog. Theo. Phys.}, 127, 2012.

\bibitem{Mand012a}
D.~Mandal and C.~Jarzynski.
\newblock {\em Proc. Natl. Acad. Sci. USA}, 109(29):11641--11645, 2012.

\end{thebibliography}


\begin{thebibliography}{1}

\bibitem{Futz_Szil29a}
L.~Szilard.
\newblock {\em Z. Phys.}, 53:840--856, 1929.

\bibitem{Smol12a}
M.~v.~Smoluchowski.
\newblock {\em Physik. Zeit.}, 13:1069--1080, 1912.

\bibitem{Smol14a}
M.~v.~Smoluchowski.
\newblock pages 89--121, Berlin, Germany, 1914. Teuber und Leipzig.

\bibitem{Feyn63a}
R.~Feynman, R.~B. Leighton, and M.~Sands.
\newblock {\em The Feynman Lectures on Physics--Volume 1}.
\newblock Addison-Wesley, Reading, Massachusetts, 1963.

\bibitem{Maye87a}
G.~Mayer-Kress and H.~Haken.
\newblock {\em Comm. Math. Phys.}, 111:63--74, 1987.

\bibitem{Crut08a}
J.~P. Crutchfield, C.~J. Ellison, and J.~R. Mahoney.
\newblock {\em Phys. Rev. Lett.}, 103(9):094101, 2009.

\bibitem{Crut08b}
C.~J. Ellison, J.~R. Mahoney, and J.~P. Crutchfield.
\newblock {\em J. Stat. Phys.}, 136(6):1005--1034, 2009.

\bibitem{Saga12a}
T.~Sagawa.
\newblock {\em Prog. Theo. Phys.}, 127, 2012.

\bibitem{Cove06aSupp}
T.~M. Cover and J.~A. Thomas.
\newblock {\em Elements of Information Theory}.
\newblock Wiley-Interscience, New York, second edition, 2006.

\end{thebibliography}

\onecolumngrid
\newpage
\begin{center}
\large{Supplementary Materials}
\end{center}

\setcounter{equation}{0}
\setcounter{figure}{0}
\setcounter{table}{0}
\setcounter{page}{1}
\makeatletter
\renewcommand{\theequation}{S\arabic{equation}}
\renewcommand{\thefigure}{S\arabic{figure}}
\renewcommand{\bibnumfmt}[1]{[S#1]}
\renewcommand{\citenumfont}[1]{S#1}

\section{Szilard Map Construction Details}

Here, we mention several details underlying the construction of $\TSzil$ and
$\TBake$. There is, in fact, a family of related maps. Partly, this comes from
the incompleteness of Szilard's presentation \citesupp{Futz_Szil29a}; partly, due to
possible, permitted variations in implementation. The following comments only
hint at these variations. A sequel will develop them more systematically.

\paragraph*{Erasure:} $\TE$ maps both the $A$ and $B$ Demon memory mesostates
back to a selected Demon memory reset mesostate. If that reset state is $A$,
then the mapping is:
\begin{align*}
\TE^A(x,y) =
\begin{cases}
  (x,y\delta) & y < \gamma ~, \\
  \left(x,\delta\gamma+\frac{y-\gamma}{1-\gamma}\gamma(1-\delta) \right)
  & y \geq \gamma ~.
\end{cases}
\end{align*}

If the reset state is $B$, then the mapping is:
\begin{align*}
\TE^B(x,y) =
\begin{cases}
  \left(x, \gamma - y + \frac{y}{\gamma}
	\left(1 - \delta (1-\gamma) \right) \right) & y < \gamma ~, \\
  \left(x, 1 - \delta(1-y) \right) & y \geq \gamma ~.
\end{cases}
\end{align*}

The cell boundary required by area preservation (probability invariance) under
map $\TE^A$ is $y^\prime = \delta \gamma$ and for map $\TE^B$, $y^\prime =
1-\delta (1-\gamma)$. The reason for this is that during the control operation
horizontal stretching by $\delta^{-1}$ or by $(1-\delta)^{-1}$ ``dilutes the
gas'' or reduces the probability density. To maintain probability invariance
(or ``gas density'') we must compensate in the erase operation by multiplying
the Demon (vertical) coordinate by $\delta$ or $1 - \delta$.

\begin{figure*}[t]
\begin{center}
\includegraphics[width=0.45\columnwidth]{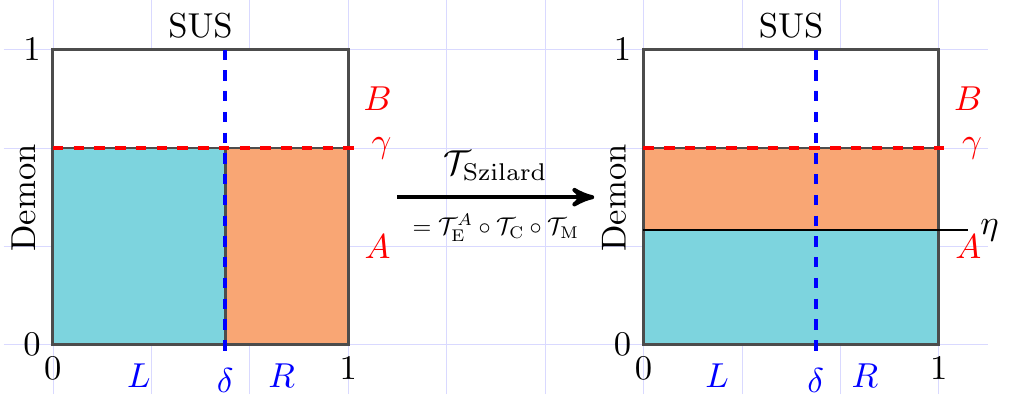}
\hspace{0.2in}
\includegraphics[width=0.45\columnwidth]{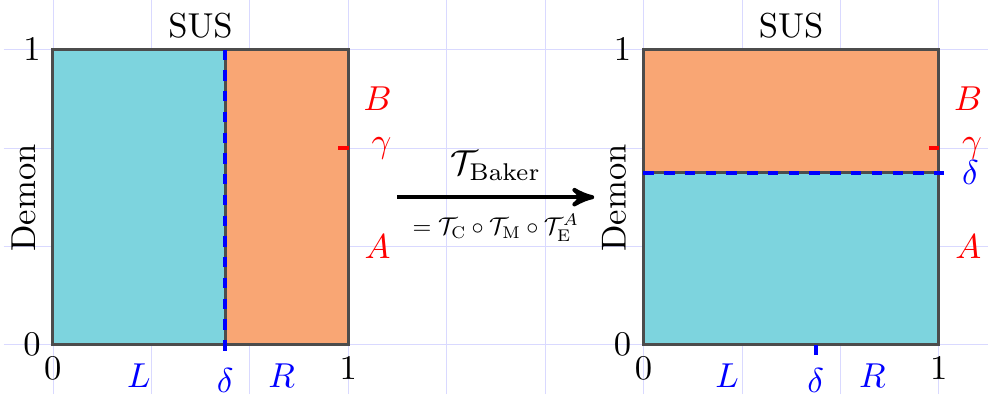}\\
(a) \hspace{3.2in} (b)
\end{center}
\caption{Dynamical Szilard Engine Composite Maps:
  (a) Szilard Map $\TSzil = \TEr^A \circ \TCon \circ \TMeas$ 
  and (b) Asymmetric Baker's Map
  $\TBake = \TCon \circ \TMeas \circ \TEr^A$.
  These assume the Demon's reset memory state is $A$.
  Areas left and right of $\delta$ are colored only as an aid to understanding
  the maps' action and to track the particle location history.
  }
\label{fig:SMapComposite}
\end{figure*}

\paragraph*{Full cycle operation:}
The overall information engine cycle, then, is the map given by the
composition of the component maps: $\TSzil = \TE^A \circ \TC \circ \TM$.
The action of $\TSzil$ is shown in Fig. \ref{fig:SMapComposite}(a), where
$\eta = \gamma \delta$. The map is area preserving, as is verified below by
calculating the determinant of its Jacobian.

\paragraph*{Initial-reset engine:}
If one first resets the engine with $\TE^A$, then we obtain the familiar
Baker's Map $\TBake = \TC \circ \TM \circ \TE^A$:
\begin{align*}
\TBake (x,y) = 
 & \begin{cases}
 (\frac{x}{\delta},\delta y)
 & x < \delta ~,\\
 \left( \frac{x-\delta}{1-\delta}, \delta + y (1-\delta) \right)
 & x \geq \delta ~.
\end{cases}
\end{align*}
It's action on the joint state space is show in Fig. \ref{fig:SMapComposite}(b).

Due to several choices in the construction of the component maps, the two
distinct $\TSzil$ maps that individually require proper initialization (reset
to $A$ or reset to $B$) can be combined into a single, more general
$\TSzilComposite$ that does not. The resulting map simply operates on what it
is given as initial conditions $(x_0,y_0) \in \mathbb{I}^2$. Mesostates in one
or the other reset memory are properly transformed. This composite map is:
\begin{align*}
\TSzilComposite (x,y) =
  & \begin{cases}
 	(\frac{x}{\delta},\delta y)
 & x < \delta, y < \gamma~,\\
 	\left(\frac{x-\delta}{1-\delta},\delta\gamma+y(1-\delta) \right)
 	& x \geq \delta, y < \gamma~, \\
 \left(\frac{x}{\delta},
 	\delta\gamma + \frac{y-\gamma}{1-\gamma} \gamma (1-\delta) \right)
	& x < \delta, y \geq \gamma~,\\
 \left(\frac{x-\delta}{1-\delta},
 	\frac{y-\gamma}{1-\gamma} \gamma\delta \right)
	&  x \geq \delta, y \geq \gamma
 ~.
\end{cases}
\end{align*}

\section{Szilard Engine Dynamical System: An Animation}

Probably the most direct way to understand the operation of Szilard's Engine is
via a continuous-time embedding of $\TSzil$. Animations can be viewed at
\url{http://csc.ucdavis.edu/~cmg/compmech/pubs/dds.htm}.

Here, we provide accompanying descriptive text that addresses several issues:
How the discrete-time $\TSzil$ is embedded in continuous time, what the
protocol cycle looks like in operation, where the instability and chaos occur
(during energy extraction), what erasure affects, and, notably, the appearance
of self-similar (fractal) structure in the joint state-space distribution.

\paragraph*{Background:} In 1929 Szilard \citesupp{Futz_Szil29a} gave Maxwell's
Demon its first logical resolution---that ``intelligence'' was not violating
the Second Law---by accounting for the Demon's own, what we now
call, information processing:
\begin{quote}
\ldots we must conclude that the intervention which establishes the coupling
between $y$ and $x$, the measurement of $x$ by $y$, must be accompanied by a
production of entropy.
\end{quote}
(Cf. Smoluchowski's ``physical'' resolution \citesupp{Smol12a,Smol14a,Feyn63a}.)

\paragraph*{Szilard Engine Operation:}
There are two ways to describe the Engine's operation: as its action on a
single particle or that on an infinite ensemble of particles via evolving a
probability density. The graphics and animations show the latter. The former is
easier to describe, initially.

Consider a single particle in a box, whose position the Demon determines by
placing a partition at $x = \delta$. Then, noting on which side (left or right)
the particle falls, the Demon lets the particle push the partition (right or
left, respectively) to extract energy ($\int_\delta^1 P~dV$ or $\int_\delta^0
P~dV$, respectively).  (See Sec. \ref{sec:engthermo} for the thermodynamic
calculations.) The entire system is held at constant temperature $T$ by a heat
bath.  Finally, the Demon's memory is reset to its original state
(mesostate $A$ or $B$).

Our construction of $\TSzil$ demonstrates that Szilard's Engine is, in fact, a
deterministic chaotic map of the unit square $\mathbb{I}^2 = [0,1] \times
[0,1]$. The construction ``symmetrizes'' the Demon and System Under Study
(SUS).  (The latter is the name for the particle-in-a-box thermodynamic
subsystem.) That is, we look at the joint state-space of both the Demon and SUS.

\paragraph*{Interpretation:} In the ensemble view, information processing
describes changes in the support and shape of the joint state-space
distributions. The thermodynamics describes the energy flow to/from the heat
bath during each protocol step.

\paragraph*{Animation:} This is a continuous-time embedding of the 2D map.
Suspending a discrete-time map in a continuous-time flow is a standard, if
somewhat under-utilized, visualization technique \citesupp{Maye87a}.

The animation shows the evolution of the Engine state-space distribution,
beginning when the Demon starts in its ``reset'' mesostate $A$. Two pieces of
the distribution are colored to correspond to the particle in the left or right
side of the partition placed at $x = \delta$. This is an aid to visually track
the particle's location and also to highlight the component maps' actions on
areas.

The animation steps through the protocol (i) Measurement (Demon memory state
and particle location come into correlation), (ii) Control (extract energy from
heat bath), and (iii) Erase (clear Demon memory to start the cycle anew).

The parameter $\gamma \in [0,1]$ represents the division between Demon memory
mesostates $A$ and $B$.

$\gamma$ and $\delta$ together let us explore the ``efficiency'' of Demon
measurements of particle location.

The energy extraction (``gas expansion'') corresponds to the state-space
stretching during the thermodynamic control step $\TC$. This is particularly
noteworthy: \emph{The instability of chaos is essential to energy extraction.}

Crucially, one sees another consequence of the chaotic dynamics: The build-up
over each cycle of the self-similar (fractal) structure of the joint
state-space distribution.

The construction allows one to make a solid connection between thermodynamics,
information processing, and chaotic dynamics. Indeed, everything can be
analytically calculated. Leaving few, if any, remaining mysteries for the
Szilard Engine, as originally conceived. In this light, Szilard's conclusion
takes on new meaning \citesupp{Futz_Szil29a}:
\begin{quote}
\ldots a simple inanimate device can achieve the same essential result as
would be achieved by the intervention of intelligent beings. We have examined
the ``biological phenomena'' of a nonliving device and have seen that it
generates exactly that quantity of entropy which is required by
thermodynamics.
\end{quote}
With this firm foundation, designing and analyzing more sophisticated
thermodynamic control systems becomes possible, including monitoring
information creation and the necessary attributes of ``intelligence''.

\section{Spectrum of Lyapunov Characteristic Exponents}

The Szilard Map Jacobian $\partial_{xy} \TSzil$ determines its local
linearization and so one can easily calculate the spectrum of Lyapunov
characteristic exponents (LCEs) for each thermodynamic step and for the overall
cycle \cite{Laso85a,Dorf99a}. This gives insight into the directions
(submanifolds) of stability (convergence) and instability (divergence). We work
with $\TBake$. There are two LCEs. We find that one is positive:
\begin{align*}
\lambda^+ & = \lim_{t \to \infty} t^{-1}
  \log_2 \big|\partial_{xy} \TBake^t (x_0,y_0) \cdot \delta \vec{x} \big|
  ~,
\end{align*}
where, since the unstable manifolds parallel the $x$-axis, the initial vector
is $\delta \vec{x} = (1,0)$. Due to ergodicity, we can calculate via:
\begin{align*}
\lambda^+ & = \int_0^1 dx
  \left[ \int_0^\delta dy ~\widehat{\rho}~ \log_2 \delta
  + \int_\delta^1 dy ~\widehat{\rho}~ \log_2 (1-\delta) \right] \\
  & = \H(\delta)
  ~.
\end{align*}
There is no dependence on initial condition $(x_0,y_0) \in \mathbb{I}^2$.
There is a companion negative LCE that monitors state space contraction.
Since the stable manifolds parallel the $y$-axis, we take the initial vector
$\delta \vec{y} = (0,1)$, finding:
\begin{align*}
\lambda^- & = \lim_{t \to \infty} t^{-1}
  \log_2 \big|\partial_{xy} \TBake^t (x_0,y_0) \cdot \delta \vec{y} \big| \\
  & = - \H(\delta)
  ~.
\end{align*}

Note that state-space area preservation is reflected in the exact balance of
instability and stability: $\lambda^+ + \lambda^- = 0$. More specifically, the
unstable manifold supports the mechanism that amplifies small fluctuations from
the heat bath to macroscopic scale and so to extractable work. The stable
manifold is the mechanism that dissipates energy into the ambient heat bath,
due to Demon memory resetting.

\section{Intrinsic Computation Calculations}

\begin{figure}[htp]
\begin{center}
\includegraphics[width=0.4\columnwidth]{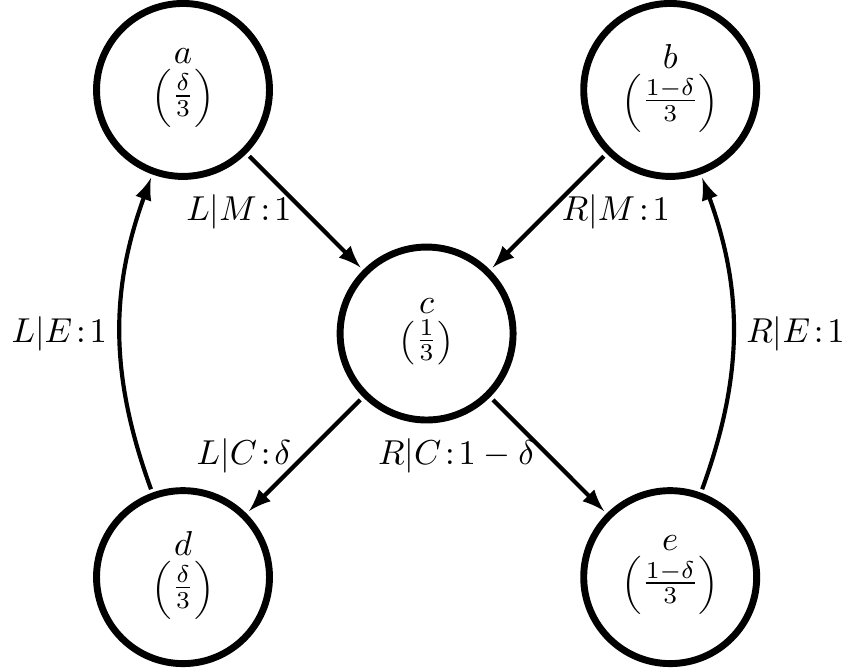}
\hspace{0.13in}
\includegraphics[width=0.3\columnwidth]{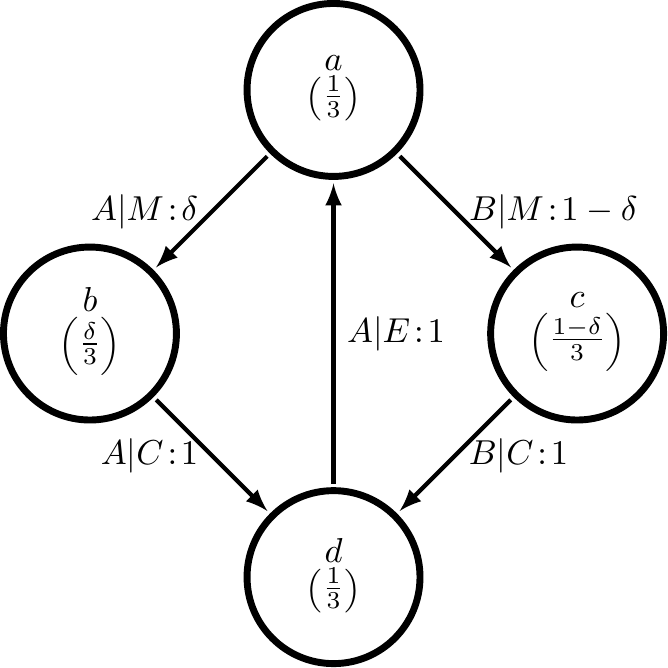} \\
{\small \hspace{0.1\columnwidth} (a) SUS \eT. \hspace{0.2\columnwidth} (b) Demon \eT.}\\
\end{center}
\caption{
\ETs\ for the symbolic dynamics of the Szilard information engine from
  the Markov partition of the Demon and SUS state spaces: (a) \ET\ for
  the thermodynamic system, giving the $\{R,L\}$-symbolic dynamics of
  particle location when the engine's \eT\ is driven by the period-$3$
  measurement protocol: $M \to C \to E \to \ldots$.  (b) \ET\ for the
  Demon's memory, with the engine similarly driven.
  \ET\ transitions $\beta|\alpha:p$ denote taking
  a transition with probability $p$ on seeing $\alpha = \{M, E, C\}$ and
  emitting symbol $\beta \in \{R,L\}$ or $\beta \in \{A,B\}$.
  Asymptotic state probabilities are given in parentheses underneath
  state names.
  }
\label{fig:SzilardEMSupp}
\end{figure}

We can project the global symbolic dynamics, captured by the \eT\ of Fig.
\ref{fig:SzilardEM}(a), onto just that for the Demon (over alphabet $\{A,B\}$)
or just that for the SUS (over alphabet $\{L,R\}$). The SUS and Demon \eTs\ are
shown in Fig. \ref{fig:SzilardEMSupp}(a) and \ref{fig:SzilardEMSupp}(b),
respectively. Comparisons are insightful.

The projections make it clear that the SUS and Demon \eTs\ are duals
of each other. When one splits into multiple causal states, the other contracts
its causal states, as expected. Branching and contraction must be balanced in a
state-space volume preserving system.

\newcommand{\cs}{\causalstate}
\newcommand{\CS}{\CausalState}

Computational mechanics analysis shows that the SUS system has
an entropy rate of:
\begin{align*}
\SUShmu & = \H[X|\CS] \\
  & = - \sum_{\cs \in \CausalStateSet^{\text{SUS}}} \Pr(\cs)
		\sum_{x \in \{L,R\}} \Pr(x|\cs) \log_2 \Pr(x|\cs) \\
  & = \H(\delta)/3
  ~,
\end{align*}
where $\CausalStateSet^{\text{SUS}}$ is the set of its causal states,
and a statistical complexity of:
\begin{align*}
\SUSCmu & = \H[\Pr(\CS)] \\
  & = - \sum_{\cs \in \CausalStateSet^{\text{SUS}}} \Pr(\cs) \Pr(\cs) \log_2 \Pr(\cs) \\
  & = \log_2 3 + 2/3 \H(\delta)
  ~.
\end{align*}
Note that the time scale here is measured in individual protocol steps.

The Demon's memory process has the same entropy rate $\Dhmu = \H(\delta)/3$,
but different
statistical complexity $\DCmu = \log_2 3 + 1/3 \H(\delta)$. For the overall
Szilard information engine (Fig. \ref{fig:SzilardEM}(a)) we find that it has
the same entropy rate $\hmu^{\text{Szilard}} = \H(\delta)/3$, but statistical
complexity $\Cmu^{\text{Szilard}} = \log_2 3 + \H(\delta)$. The analysis shows
us that the SUS and the Demon subsystems are not independent, however. The
joint $\Cmu^{\text{Szilard}}$ is substantially less than the sum of that of the
two subsystems: $\Cmu^{\text{Szilard}} = \SUSCmu + \DCmu - \log_2 3$. Not
surprisingly, the two subsystems are redundant in their operation. The $\log_2
3$ bits of memory reflect the synchronization of SUS and Demon during the
period-$3$ control protocol, which is accounted for only once in the overall
Engine $\Cmu^{\text{Szilard}}$.

An information engine is a control system and its effectiveness depends on the
controller (Demon) knowing the state of the thermodynamic subsystem (SUS). So,
how predictable is the SUS from the Demon's perspective? And, for that matter,
since our construction is symmetrized, the Demon from the viewpoint of the SUS?
The answers start with the information engine's past-future mutual information,
the \emph{excess entropy}: $\EE^{\text{Szilard}} = \I[\IEPast;\IEFuture]$,
where $\IEPast = \ldots \IESymbol_{-3} \IESymbol_{-2} \IESymbol_{-1}$ is the
past and $\IEFuture = \IESymbol_0 \IESymbol_1 \IESymbol_2 \ldots$ is the
future of the joint process over random variable $\IESymbol_t \in \{A,B\} \otimes \{R,L\}$.
$\EE^{\text{Szilard}}$ measures the amount of future information predictable
from the past. We can also ask about the predictability of the individual SUS
and Demon subprocesses using $\SUSEE$ and $\DEE$, respectively, defined
similarly. We see that the machines in Fig. \ref{fig:SzilardEMSupp} are
counifilar \citesupp{Crut08a,Crut08b} and so $\EE^{\text{Szilard}} =
\Cmu^{\text{Szilard}}$, $\DEE = \DCmu$, and $\SUSEE = \SUSCmu$
\citesupp{Crut08b}. (Section \ref{sec:SuppMI} calculates the inter-subsystem
information flows.)

\section{Engine Thermodynamics}
\label{sec:engthermo}

The minimum energy cost of the measurement, control, and erasure protocol steps can
be calculated by treating the Szilard Engine as a box of ideal gas. (Sequels
address more realistic models of the working fluid.) Each of the
transitions shown in Fig. \ref{fig:SMap} is achieved by isothermal sliding of
barriers, so the work done is exactly calculable by integrating the force on
these barriers. The work done on this system is equal to the heat dissipated,
since the thermal energy of the particle does not change for an isothermal
process.

For each component-map thermodynamic transformation, the work done on the
system is given by $W=-\int P dV$, where $P = k_B T / V$ for a single
particle. Thus:
\begin{align*}
W = -\int_{V_0}^{V_1}\frac{k_B T}{V}dV=k_B T \ln \frac{V_0}{V_1}
  ~,
\end{align*}
where $V_0$ is the initial volume of the region being manipulated and $V_1$ is
the final volume.

Assume the Demon resets to mesostate $A$.

To calculate the thermodynamic cost of measurement $\langle \QM \rangle$, note that there are two regions
where the particle can exist. The particle starts in a uniform distribution
over the $L \otimes A$ and $R \otimes A$ regions. The probability of being in
each of those regions is proportional to the the region's volume. Assuming the
box has volume $V$, the size of region $L \otimes A$ is $\delta \gamma
V$ and the size of region $R \otimes A$ is $(1-\delta) \gamma V$.  Thus,
before measurement, the probability of being in $R \otimes A$ is $(1-\delta)$.

The region $L \otimes A$ is unchanged under the measurement map $\TM$, meaning that there is no
work done in the case where the particle is on the left.  However, if the
particle is on the right, volume is being moved to the $R \otimes B$ region, so
we have a change in volume.  The resulting average work done is the average
dissipated heat:
\begin{align*}
\langle \QM \rangle & = \langle W_\text{measure} \rangle \\
  & = \Pr(L \otimes A) \times 0
  + \Pr(R \otimes A) \times -\int_{V_{R \otimes A}}^{V_{R \otimes B}} P dV \\
  & = 0 + (1-\delta) k_B T \ln \frac{V_{R \otimes A}}{V_{R \otimes B}} \\
  & = -k_B T (1-\delta) \ln\frac{1-\gamma}{ \gamma}
  ~.
\end{align*}

Before the thermodynamic control transformation $\TC$, the particle has probability
$\delta$ of being in the $L \otimes A$ region and probability $1-\delta$ of
being in the $R \otimes B$ region.  If the particle is in the $L \otimes A$
region, then after control the particle occupies the whole $A$ region, and if
it was in $R \otimes B$, then it occupies the whole $B$ region.  According to
the change in volume, the average heat dissipated is:
\begin{align*}
\langle \QC \rangle
  & = \langle W_\text{control} \rangle \\
  & = -\Pr(L \otimes A) \int_{V_{L \otimes A}}^{V_{L \otimes A}+V_{R \otimes A}}\frac{k_B T}{V}dV
  - \Pr(R \otimes B) \int_{V_{R \otimes B}}^{V_{R \otimes B}+V_{L \otimes B}}\frac{k_B T}{V}dV \\
  & =- \delta k_B T \ln \frac{V \gamma}{V \delta \gamma}
  - (1-\delta) k_B T \ln \frac{V (1-\gamma)}{V (1-\delta) (1-\gamma)} \\
  & = -k_B T ( -\delta \ln \delta - (1-\delta) \ln (1-\delta)) \\
  & = -k_B T \ln 2 \, \H(\delta)
  ~.
\end{align*}

Before the erasure transformation $\TE$, the particle has probability $\delta$ of
being distributed uniformly over the $A$ region and probability $1-\delta$ of
being distributed uniformly over the $B$ region. The $B$ region is compressed
into the region between $\eta = \gamma \delta$ and $\gamma$ on the vertical demon
axis, which has volume $\gamma(1-\delta)V$. The $A$ region is compressed
between $0$ and $\eta$ on the Demon axis, which has volume $\gamma \delta V$.
The associated heat dissipation is, then:
\begin{align*}
\langle \QE \rangle &= \langle W_\text{erase} \rangle \\
  & = - \Pr(B) \int_{(1-\gamma)V}^{\gamma(1-\delta)V}\frac{k_B T}{V}dV
  - \Pr(A)\int_{\gamma V}^{\gamma \delta V}\frac{k_B T}{V}dV \\
  & = - k_B T ((1-\delta)\ln \frac{\gamma(1-\delta)}{1-\gamma}
  + \delta \ln \delta) \\
  & = k_B T \ln 2 \, \H(\delta)+k_B T (1-\delta) \ln \frac{1-\gamma}{\gamma}
  ~.
\end{align*}
After erasure, the particle is once again uniformly distributed over the $A$ region.

We close by commenting on the main text's mention of the Szilard Engine's optimality. We note that the results above match those in Ref. \citesupp{Saga12a} for $\delta = 1/2$. Moreover, they achieve the bounds described there for all $\delta$ and $\gamma$. Specifically, for $\delta = 1/2$, our results match Eq. (7.23) for erasure and Eq. (7.40) for measurement.
And, our results achieve the lower bounds given there in Eqs. (7.41),
(7.43), and (7.44), since $I_{QC} = H(\delta)$. Our development, though, places
measurement, control, and erasure on the same footing, allowing us to make
the same statements about each, which is that we quantify how much heat
they dissipate.

\section{Monitoring Correlation and Coordination During the Control Protocol}
\label{sec:SuppMI}

There are many different correlations potentially relevant to Demon functioning
and to monitoring its interaction with the SUS. Let $Y$ be the random variable
for the Demon's memory $\{A, B\}$. Let $X$ be the random variable for the SUS's
particle locations $\{L,R\}$. Let $S^{Y}$ be that for Demon's causal states
(Fig. \ref{fig:SzilardEMSupp}(b)) and $S^{X}$ that for SUS's causal states (Fig. \ref{fig:SzilardEMSupp}(a)).

The first mutual information we consider is the asymptotic communication rate
between the Demon and the SUS:
\begin{equation}
\lim_{\ell \rightarrow \infty} \frac{\I[X_{0:\ell};Y_{0:\ell}]}{\ell}
  ~,
\end{equation}
where the random variable blocks are $X_{0:\ell} = X_0 X_1 \ldots X_{\ell-1}$
and $Y_{0:\ell} = Y_0 Y_1 \ldots Y_{\ell-1}$.
This can be evaluated by breaking it into components:
\begin{align*}
\lim_{\ell \rightarrow \infty} \frac{\I[X_{0:\ell};Y_{0:\ell}]}{\ell}
   & = \lim_{\ell \rightarrow \infty}\frac{1}{\ell}
 \left( \H[X_{0:\ell}]+\H[Y_{0:\ell}]-\H[X_{0:\ell},Y_{0:\ell}] \right) \\
  & = h_\mu[\overleftrightarrow{X}] + h_\mu[\overleftrightarrow{Y}]
    - h_\mu[\overleftrightarrow{X},\overleftrightarrow{Y}]\\
  & = \H(\delta)/3 + \H(\delta)/3 - \H(\delta)/3 \\
  & = \H(\delta)/3
  ~.
\end{align*}
Note that the Shannon (base 2) binary entropy function $\H(\delta)$
\citesupp{Cove06aSupp}
features prominently in the expressions for heat dissipation. However, the
above shows that it often signifies an amount of shared or mutual information,
not just a degree of Shannon information uncertainty. These two interpretations
are rather distinct.

This should be compared to single-symbol (length-$1$ block) mutual information
between the Demon and SUS:
\begin{align}
\I[X_0;Y_0] & = \frac{\delta(2+\delta)}{3}
   \log_2 \frac{2+\delta}{1+\delta}
   + \frac{(1-\delta)(1+\delta)}{3}\log_2 \frac{1+\delta}{2+\delta} \\
  & \quad - \frac{\delta(1-\delta)}{3}
   + \frac{(1-\delta)(2-\delta)}{3}\log_2\frac{2-\delta}{2-2\delta}
  ~.
\end{align}
Note that this differs from the asymptotic rate above and so the single-symbol
quantity misestimates the degree of Demon-SUS correlation.

To measure the degree of internal coordination between the SUS and the Demon,
consider now the single-symbol mutual information between the SUS and Demon
causal states. First, we find the stationary distribution over the causal
states in the joint process (Fig. \ref{fig:SzilardEM}(a)) and label the joint
causal states with the corresponding causal states of the SUS and Demon. The
resulting single-state mutual information is:
\begin{align*}
\I[S^{X}_0;S^Y_0] = \log_2 3
  ~.
\end{align*}
This tells us the correlation between SUS and Demon causal states is limited to their synchronization within the three step cycle of measurement, control, and erasure.

We can similarly calculate the correlation rate between the Demon and SUS
states. For this we need the \eTs\ for the joint, Demon, and SUS processes in
Figs. \ref{fig:SzilardEM}(a), \ref{fig:SzilardEMSupp}(a), and
\ref{fig:SzilardEMSupp}(b), respectively. It turns out that the entropy rate
for all three machines is also $\H(\delta)/3$ as well, so:
\begin{align*}
\lim_{\ell \rightarrow \infty}
  \frac{\I[S^X_{0:\ell};S^Y_{0:\ell}]}{\ell}
  & = \H(\delta)/3+\H(\delta)/3-\H(\delta)/3 \\
  & = \H(\delta)/3
  ~.
\end{align*}
This equals the communication rate between the Demon and SUS symbolic processes
above, indicating that there is little ``hidden'' internal state information
compared to the symbolic processes. The Szilard Engine is not a cryptic process
\citesupp{Crut08a}.

Last, adopting the transducer perspective, we calculate the single-symbol
conditional mutual information between the Demon and SUS variables given the
three possible control protocol inputs. These monitor the dependence of
correlation during the individual protocol steps. Let our input control
variable be $Z$ over alphabet $\{M, C, E\}$. For instance, during the
measurement step, we directly calculate:
\begin{align*}
\I[X_0;Y_0|Z_0=M] & =
  \sum_{y \in \{A,B\}}
  \sum_{x \in \{L,R\}}
  \Pr(X_0=x,Y_0=y|Z_0=M) \log_2
  \left( \frac{\Pr(X_0=x,Y_0=y|Z_0=M)}{\Pr(X_0=x|Z_0=M)\Pr(Y_0=y|Z_0=M)}\right)
  \\
  & = \H(\delta)
  ~.
\end{align*}
Similarly, we calculate that during erasure there is no correlation:
\begin{align*}
\I[X_0;Y_0|Z_0=E]=0
  ~,
\end{align*}
and that during thermodynamic control the same holds true:
\begin{align*}
\I[X_0;Y_0|Z_0=C]=0
  ~.
\end{align*}
Thus, the single-symbol correlation between Demon and SUS comes from
measurement. There is also a connection between these quantities and the work
extracted during control. The work extracted in control is $\braket{W}=k_B T
\H(\delta) \ln 2$. That is, the Engine uses the decrease in correlation of
mesostates between measurement and control to extract work from the thermal
bath.

Note that these calculations for the Szilard Engine are straightforward. We
detail them here to show the individual analyses, uncluttered by distracting
calculational moves, that one must do for more complex information engines. In
the latter cases, especially when the thermodynamical system is nonhyperbolic,
the calculations can be quite challenging. Sequels illustrate.

\vspace{-0.2in}

\bibliographysupp{chaos}

\end{document}